\documentclass[twocolumn,showpacs,preprintnumbers,amsmath,amssymb]{revtex4}

\usepackage[unicode]{hyperref}
\usepackage[caption=false]{subfig}
\usepackage{graphicx}
\graphicspath{ {./images/} }
\usepackage{dcolumn}
\usepackage{bm}
\usepackage[caption=false]{subfig}

\begin{document}

\title{Redefinition of site percolation in light of entropy and the second law of thermodynamics  
}%

\author{ M. S. Rahman and M. K. Hassan 
}%
\date{\today}%

\affiliation{
University of Dhaka, Department of Physics, Theoretical Physics Group, Dhaka 1000, Bangladesh \\
}

\begin{abstract}
In this article, we revisit random site and bond percolation in square lattice focusing primarily 
on the behavior of entropy and order parameter. In the case of traditional site percolation, we find that 
both the quantities are zero at $p=0$ revealing that the system is in the perfectly ordered and in the disordered
state at the same time. Moreover, we find that entropy with $1-p$, which is the 
equivalent counterpart of temperature, first increases and then decreases again 
but we know that entropy with temperature cannot decrease. However, bond percolation does not 
suffer from either of these two problems. To overcome this we propose a new definition for site percolation 
where we occupy sites to connect bonds and we measure cluster size by the number of bonds 
connected by occupied sites. This resolves all the problems without affecting any of the existing known results.
\end{abstract}

\pacs{61.43.Hv, 64.60.Ht, 68.03.Fg, 82.70.Dd}

\maketitle

\section{Introduction}

Percolation has been studied extensively in statistical physics due to the simplicity of its 
definition and the versatility of its application in seemingly disparate complex systems. 
The reason for its simplicity is that it requires neither quantum nor
many particle interaction effects and yet it can describe phase transition
and critical phenomena \cite{ref.Stauffer, ref.saberi}. 
To define percolation, we first need to choose a skeleton. It can either be a graph 
embedded in an infinite dimensional abstract space or it can be a lattice embedded in 
an Euclidean space. In either case they always 
consist of nodes or sites and links or bonds.  Percolation 
is known as site or bond type depending on whether we occupy sites or bonds respectively.
In the case of random bond percolation, we assume that all the labelled
bonds are initially frozen. The rule  is then  to choose 
one frozen bond at each step randomly with uniform probability and occupy it. We continue the process one by one till the 
occupation probability $p$, fraction of the total bonds being occupied, reaches to unity. 
At $p=0$ each site is a cluster of its own size and as we tune $p$
we observe that clusters, a group of sites connected by occupied bonds, are continuously
formed and grown on the average. In the process there 
comes a critical point $p_c$ above which a spanning cluster $s_{{\rm span}}$ that 
spans across the entire lattice emerges. Interestingly, its relative size $P=s_{{\rm span}}/N$ varies 
with $p$ such that $P>0$ only at $p>p_c$ and $P=0$ at $p\leq p_c$. Moreover, it grows above $p_c$
following  a power-law $P\sim (p-p_c)^\beta$. This is reminiscent of the order parameter of continuous phase transition
and hence $P$ is regarded as the order parameter for percolation \cite{ref.Stanley, ref.Binney}.

Despite more than five decades long history of extensive studies we still have many unresolved issues in percolation.  For instance, we know that the
order parameter in general measures the extent of order. However, what order really
means here is not known in percolation. Finding a way to associate
order with the relative size of spanning cluster is still an issue. It has been shown 
that $P=0$ in the entire regime $0<p\leq p_c$ at least in the thermodynamic limit and hence 
we must find a way to regard it as the disordered phase. We therefore need another
quantity that can quantify the degree of disorder where $P=0$. 
The obvious choice is entropy $H$. In fact, no model for phase transition is complete without it since, 
like order parameter, entropy is also used as a litmus test to define
the order of transition. Despite being such an important quantity, its definition 
remained elusive for five decades as the first article on entropy appeared in 1999 \cite{ref.tsang}. After that
we find only two more articles \cite{ref.tsang_1, ref.Vieira} prior to our recent works since 2017 \cite{ref.hassan_didar, ref.hassan_sabbir}. 
One of the physically acceptable requirements of $H$ is that when $H$ is minimally low 
$P$ cannot be minimally low too as this would imply that the system is in the perfectly 
ordered and disordered at the same time. 
On the other hand, at and near the critical 
point both the quantities undergo an abrupt change revealing that the transition
is accompanied by symmetry breaking, as it also means an order-disorder transition. However, surprisingly this has never been an issue in percolation theory.
Besides its relevance to phase transition, the percolation model 
has also been applied to a wide variety of natural and social phenomena such as the spread of
disease in a population \cite{ref.Murray}, flow of fluid through porous media \cite{ref.fluids}, 
conductor-insulator composite materials \cite{ref.McLachlan}, resilience of systems \cite{ref.barabasi_1, ref.pastor}, 
dilute magnets  \cite{ref.Bergqvist}, the formation of public opinion \cite{ref.Watts, ref.Shao, ref.opinon_1}
and spread of biological and computer viruses leading to epidemic \cite{ref.Newman_virus, ref.Moore_virus}.

In this article, we revisit the random bond and site percolation in the square lattice focusing 
 primarily on entropy and order parameter. For bond percolation,
we find that entropy is maximum where order parameter is minimum and vice versa as expected. 
However, this is not the case for the traditional site percolation as 
we find that initially both entropy and order parameter are equal to zero which 
cannot be the case since it means that the system is in ordered and disordered state at the same time.
Moreover, we have identified that $1-p$, the fraction of the unoccupied site/bond, as the equivalent counterpart
of temperature. Thus, entropy must always increase with $1-p$ and at most remain the same but must not decrease at any stage.
This is, however, not the case for traditional site percolation instead we find that entropy first increases from zero to its maximum value and
then decreases to zero again. It violates the second law of thermodynamics which
states that entropy of an isolated system should always increase and can never decrease again.
It warrants redefinition of site percolation. 
We therefore propose a new definition for site percolation 
where it is assumed that bonds are already present in the system and we
occupy sites one by one to connect the bonds.  We then measure the cluster size in terms of the number 
of bonds connected by occupied sites. We then find that entropy for
redefined site percolation is always increases with $1-p$ and it is consistent with the corresponding order parameter.
We then argue that the opposing nature of order parameter and entropy suggest that percolation
transition is accompanied by symmetry breaking like ferromagnetic transition. 
 Besides, we reproduce all the known results for redefined site percolation which confirms that 
random bond and re-defined site percolation belong to the same universality class. Furthermore,
despite the difference between the old and new definition of site percolation they still give
the same critical point $p_c$. However, the same is not true for site and bond percolation 
as their $p_c$ values are different albeit they belong to the same universality class.

The rest of the articles is organized as follows: In Sec. II, we briefly discuss the Newman-Ziff 
(NZ) algorithm
for simulating the percolation model is briefly described. Alongside, we also discussed the 
idea of convolution. 
is introduced and the algorithm . 
Sec. III, contains general discussions about entropy and order parameter. 
Inconsistencies of entropy and order parameter  according to old definition of site 
percolation are  presented in Sec. IV. 
The site percolation is re-defined in Sec. V to resolve the inconsistencies. 
In Sec. VI we have shown that much known site-bond universality
is still valid. The results 
are discussed and conclusions drawn in Sec. VII.

\section{Newman-Ziff (NZ) algorithm}

To study random percolation, we use Newman-Ziff (NZ) algorithm as 
it helps calculating various observable quantities over the entire range of $p$ in every 
realization instead of measuring them for a fixed probability $p$ in each realization \cite{ref.Ziff}.
On the other hand, in classical Hoshen-Kopelman (HK) we can only measure an observable quantity
for a given $p$ in every realization and this is why NZ is more efficient than HK \cite{ref.hoshen}.
To illustrate the idea, we consider the case of bond percolation first. 
According to the NZ algorithm, all the labelled bonds $i=1,2,3,..., M$ 
are first randomized and then arranged in an order 
in which they will be occupied. Note that the number of bonds with periodic
boundary condition is $M=2L^2$. In this way we 
can create percolation states consisting of $n+1$ occupied bonds
simply by occupying one more bond to its immediate past state consisting of $n$ occupied 
bonds. Initially, there are $N=L^2$ clusters of size one.
Occupying the first bond  means formation of a cluster of size two. However, 
as we keep occupying thereafter, average cluster size keeps growing 
at the expense of decreasing cluster number. Interestingly, all the observables in percolation, this
way or another, are related to cluster size and hence a proper definition of cluster becomes crucial.
One of the advantages of the NZ algorithm is that we calculate an observable, say $X_n$, as 
a function of the number of occupied bonds (sites) $n$ and use the resulting data in the convolution relation
\begin{equation}
\label{eq:convolution}
X(p)=\sum_{n=1}^N  \left( \begin{array}{c}
N \\ n \end{array}\right )p^n(1-p)^{N-n} X_n,
\end{equation}
to obtain $X(p)$ for any value of $p$. 
The appropriate weight factor for each $n$ at a given $p$ is $\sum_{n=1}^N p^n(1-p)^{N-n}$ \cite{ref.Ziff}. 
The convolution relation takes care of that weight factor and hence helps obtaining a smooth curve for $X(p)$.

\section{Entropy and Order Parameter}

The two most important quantities of interests in the theory of phase transition and critical phenomena 
are the entropy $H$ and the order parameter $P$. The reason is that they are the ones
which define the order of transition. In the first order or discontinuous phase transition,
 entropy must suffer a jump or discontinuity at the critical point which is 
why first order transition
requires latent heat. Similarly, the order parameter too must suffer a jump or discontinuity at
the critical point and that is why new and old 
phase can coexist at the same time in the first order transition. Besides, they are also used as a litmus test to check
whether the transition is accompanied by symmetry breaking or not. In the case of symmetry 
breaking,
the system undergoes a transition from the disordered state, which is characterized by maximally high $H$ and $P=0$, to the ordered state, which is characterized by non-zero
$P$ and minimally low $H$. 
Such transition happens with an abrupt or sudden
change in $P$ and $H$ but without gap or discontinuity at $p_c$. 
Percolation being a probabilistic model for phase transition, there is
absolutely no room for considering thermal entropy. To this end, the 
best candidate is definitely the Shannon entropy  
\begin{equation}
\label{eq:shannon_entropy}
H(p)=-K\sum_i^m \mu_i\log \mu_i,
\end{equation} 
provided we use a suitable probability for $\mu_i$ that gives entropy
consistent with the order parameter and the second law of thermodynamics \cite{ref.shannon}. Recently, we have shown
that the cluster picking probability (CPP) for $\mu_i(p)$, that a site picked at random belongs to 
labelled cluster $i$ at occupation probability $p$, is the appropriate
quantity for measuring entropy for percolation \cite{ref.hassan_didar, ref.hassan_sabbir}.

On the other hand, order parameter for bond percolation is the strength $P(p,L)$ of the spanning cluster for system size $L$ is defined as 
\begin{equation}
\label{eq:pp1}
P={{{\rm Number \ of \ sites \ in \ the \ spanning \ cluster}}\over{{\rm Total \ 
number \ of \ sites }}}.
\end{equation}
Essentially, it describes the probability that a site 
picked at random belongs to the spanning cluster at occupation probability $p$ for system size $L^2$ .
It has been found that in the limit $L\rightarrow \infty$ the probability $P(p,L)=0$ 
for $p\leq p_c$ and it reaches to its maximum value $P(p,L)=1$ 
following a power-law $P\sim (p-p_c)^\beta$ near but above $p_c$. This is indeed reminiscent 
of the order parameter of the continuous thermal phase transition like magnetization
during the ferromagnetic transition.  It is noteworthy to mention that
 for finite system size $P(p,L)$ has a non-zero
value even at $p<p_c$. However, as we increase the linear size $L$ of the system it always shows a 
clear sign of becoming zero at a higher value towards $p_c$. There exists yet another definition of $P$ where 
we can use the size of the largest cluster instead of the spanning cluster. 
However, both definitions give exactly the same qualitative results in the thermodynamic limit.

\begin{figure}

\centering

\subfloat[]
{
\includegraphics[height=3.6 cm, width=4.0 cm, clip=true]
{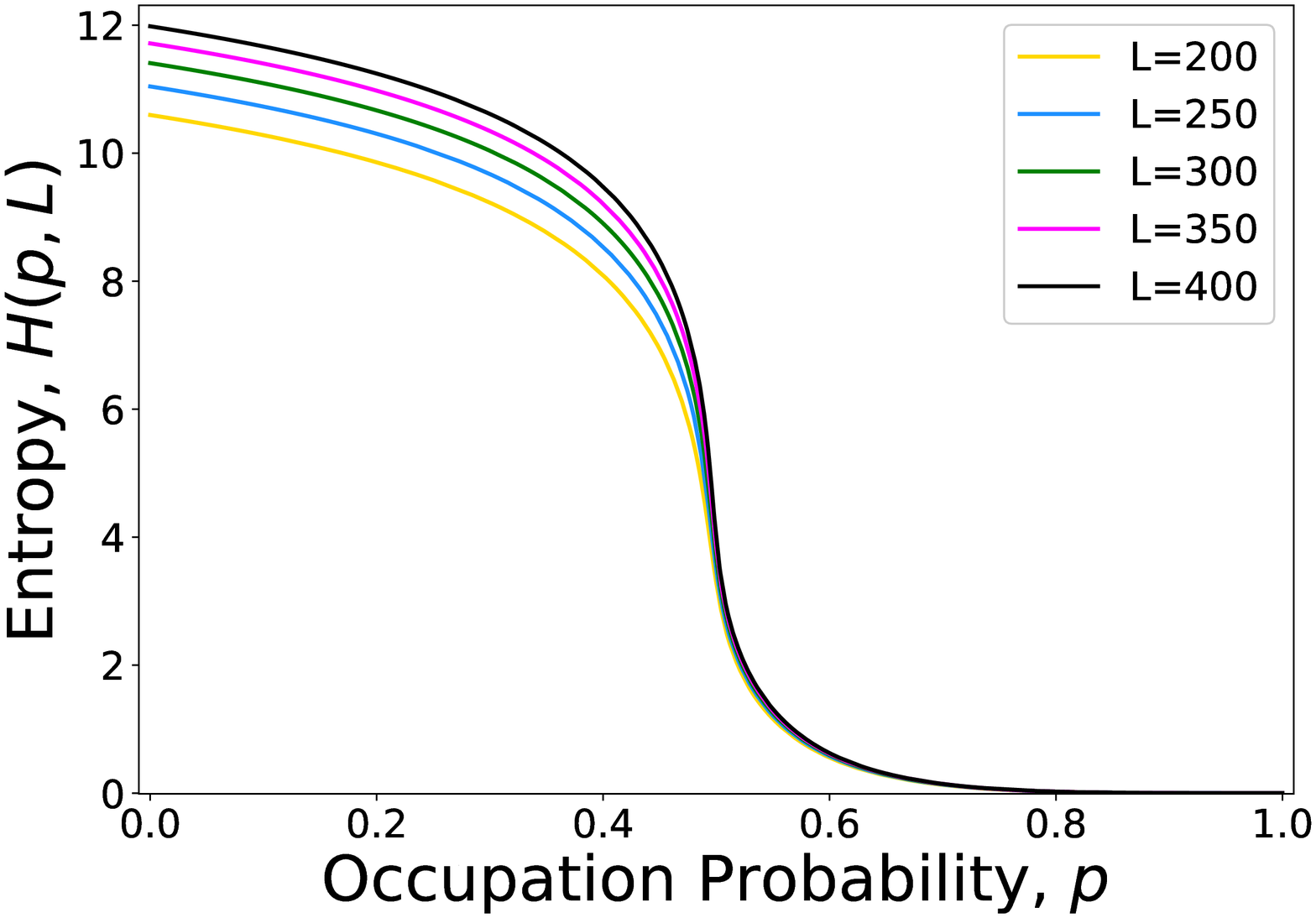}
\label{fig:1a}
}
\subfloat[]
{
\includegraphics[height=3.6 cm, width=4.0 cm, clip=true]
{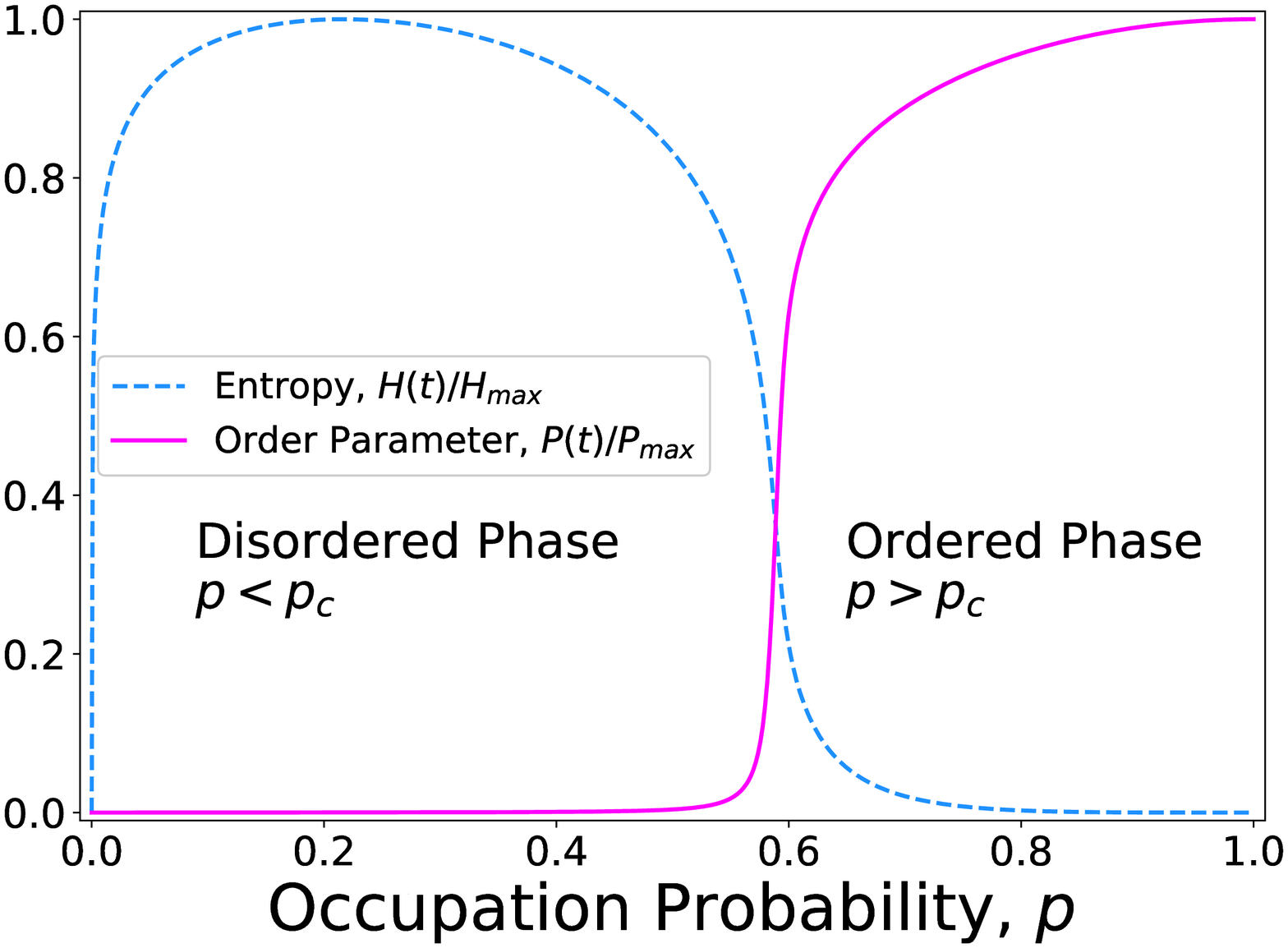}
\label{fig:1b}
}

\caption{(a) Entropy $H$ versus $p$ for bond percolation. (b) Entropy $H/H_{{\rm max}}$ (dashed blue) and order 
parameter $P/P_{{\rm max}}$ (purple) for traditional site percolation. 
} 

\label{fig:1ab}
\end{figure}

\section{Problem with existing site percolation}

We first measure entropy for random bond percolation
where initially every site is a cluster of its own size. As 
we keep occupying or reactivating frozen bonds, clusters are continuously formed and
their sizes on average are grown. Consider that at an arbitrary step of the process
there are $m$ distinct, disjoint, and indivisible labelled clusters $i=1,2,...,m$ 
whose sizes are $s_1,s_2,....,s_m$ respectively. We can therefore define 
CPP as $\mu_i=s_i/\sum_i s_i$,
that a site picked at random belongs to the labelled cluster $i$, which is naturally normalized  
$\sum_j s_j=N$ \cite{ref.hassan_didar}. Note that for convenience we choose 
$K=1$ in Eq. (\ref{eq:shannon_entropy}) since it merely amounts to a choice of a unit of measure of entropy. Clearly, at $p=0$ we have 
$\mu_i=1/N$ for all the sites $i=1,2,...,N=L^2$ which is exactly like the state 
of the isolated ideal gas where all the allowed microstates are equally likely.
Clearly, the entropy is maximum $H=\log N$ at $p=0$ revealing that we are in a state of 
maximum uncertainty just like the state of the isolated ideal gas.  
On the other hand, as we go to the other extreme at $p=1$, we find that all the 
sites belong to one cluster that makes  $\mu_1=1$. It implies
that entropy is zero at $p=1$ and hence we are in a state of zero uncertainty just like the perfectly 
ordered crystal structure. In order to see how entropy interpolates between $p=0$ and $p=1$,
we use CPP in Eq. (\ref{eq:shannon_entropy}) and the resulting entropy is shown
in  Fig. (\ref{fig:1a}) as a function of $p$ for different system sizes.

To see how entropy for random site percolation 
differs from that of the bond type, we now measure entropy for traditional definition of site percolation. In this case,
 it is assumed that initially all the sites are frozen or empty.
The process starts with the occupation of sites one by one at random and at the same time measure
the cluster size exactly like in the bond percolation.
It means that initially CPP does not exist and after the occupation of the first site we have $\mu=1$ 
and hence entropy $H=0$ at $p=1/N$ which is essentially zero
in the limit $N\rightarrow \infty$.
As we further occupy sites, we observe a sharp rise in the entropy to its maximum value, 
see Fig. (\ref{fig:1b}), which happens near $p=0.2$. Thereafter it decreases 
with $p$ qualitatively in the same way as in the case of random bond type percolation. To check
whether the behavior of entropy and order parameter are consistent or not
we have plotted both in the same figure as shown in Fig. (\ref{fig:1b}) distinguished by blue and orange colors respectively. It 
suggests that at $p\approx 0$ entropy and order parameter  both 
are zero which cannot be the case since that would mean initially the
system is perfectly ordered and disordered at the same time.

We all know from the thermodynamics
that entropy must always either increase or at best may remain constant
with temperature. On the other hand, the   
order parameter must always decrease which eventually becomes zero at the critical 
point with temperature. If we now plot entropy as a 
function $1-p$ instead of $p$ for bond percolation
we find that the entropy and order parameter both behaves exactly like they should do
 in thermal phase transition and the behaviour is consistent with the second law of thermodynamics.
The same behavior is also expected for traditional site percolation but clearly it is not the case.
These two features thus warrant re-definition of site percolation.

It is noteworthy to mention that phase transition always involves change in entropy. No model
or theory of phase transition is complete without the idea of entropy since
the nature of change in entropy defines the order of phase transition like order parameter.
Percolation has the history of extensive studies for more than $60$ years
as a paradigmatic model for continuous or second order phase transition. Yet till 2014
only two groups namely
Tsang {\it et al.} and Vieira {\it et al.} investigated entropy for percolation. Both the groups
used Eq. (\ref{eq:shannon_entropy}) but not the same expression for $\mu_i$. 
For instance, Tsang {\it et al.} used $w_s$, the probability that  
a site picked at random belongs to a cluster which contain exactly $s$ sites, to 
measure entropy for percolation. On the other hand, Vieira {\it et al.} used $n_s$, 
number of cluster of size $s$ per site,  to measure entropy $H$. In either case, they found
entropy is equal to zero at $p=0$ where order parameter is also equal to zero.
Thus, their entropy too suffers the same problem as our entropy for site percolation suffers.

\begin{figure}

\centering

\subfloat[]
{
\includegraphics[height=3.6 cm, width=4.0 cm, clip=true]
{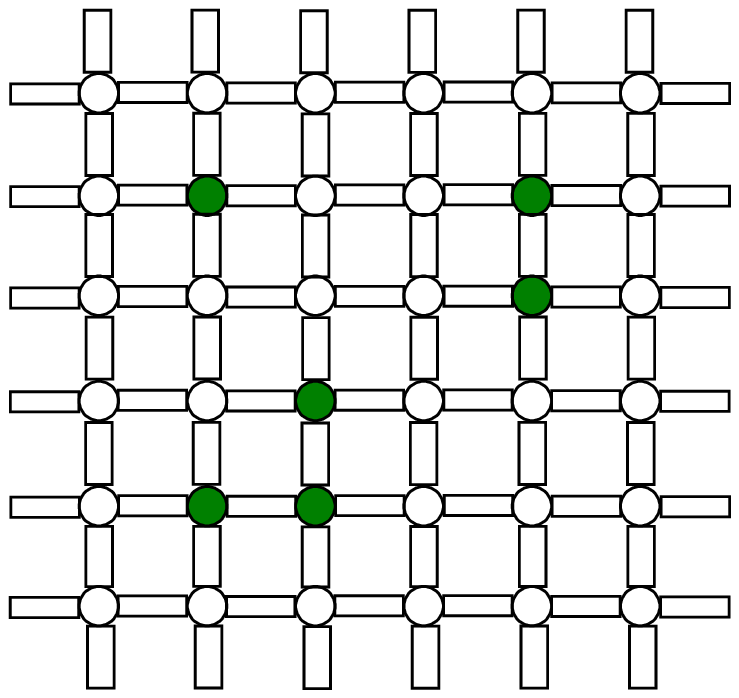}
\label{fig:image_a}
}
\subfloat[]
{
\includegraphics[height=3.6 cm, width=4.0 cm, clip=true]
{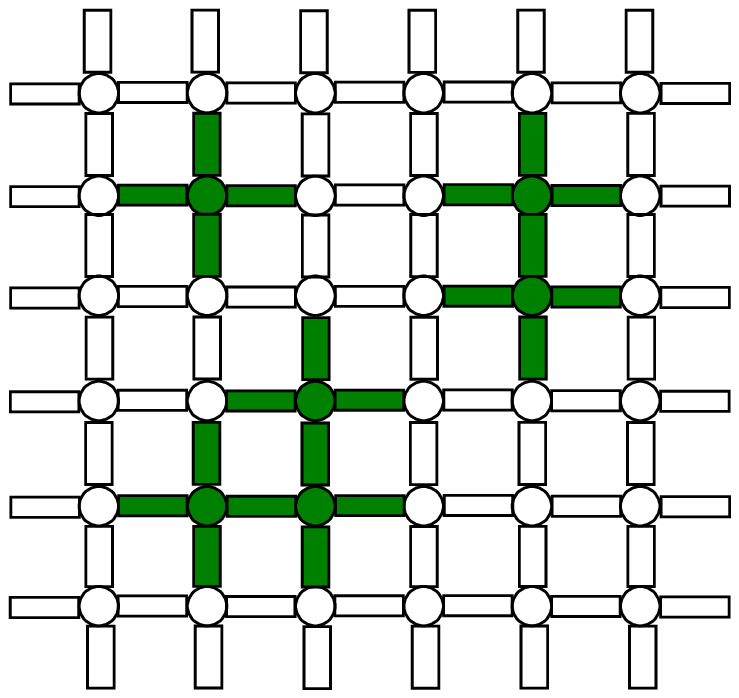}
\label{fig:image_b}
}


\caption{Schematic illustration of (a) old and (b) new definition of site percolation on square lattice. 
We assume that process starts with isolated bonds (thick black 
lines)  but sites are empty (white circles).
} 

\label{fig:image_2}
\end{figure}

\section{Site percolation re-defined}

The questions is: How can we resolve the problem with the existing definition of site percolation?
Recall the definition of bond percolation where we occupy bonds to connect the already 
existing sites in the system and measure cluster size by the number of contiguous sites connected by occupying
bonds. We already know that it is consistent with the second law of thermodynamics
and the nature of the order parameter. Using the spirit of the bond percolation
we can define the site percolation as follows. We assume that the bonds are already there 
in the system as an isolated entity and the occupation of sites connect these isolated bonds to form clusters
of bond. How do the old and new definitions differ? 
According to the old definition, occupation of one, two and three isolated empty 
sites forms a cluster of
size one, two and three as shown in  Fig. (\ref{fig:image_a}) by the green color. 
On the other hand,
according to the redefined site percolation occupation of one, two and three empty
sites forms a cluster of size four, seven and ten respectively which are shown in Fig. (\ref{fig:image_b}).  
In the case of bond percolation, on the other
hand, occupation of one isolated bond forms a cluster of size two, two consecutive isolated bonds
forms cluster of size three, three consecutive isolated bonds form cluster of size four and so on. 
Thus all three definitions are clearly different.

\begin{figure}

\centering

\subfloat[]
{
\includegraphics[height=3.6 cm, width=4.0 cm, clip=true]
{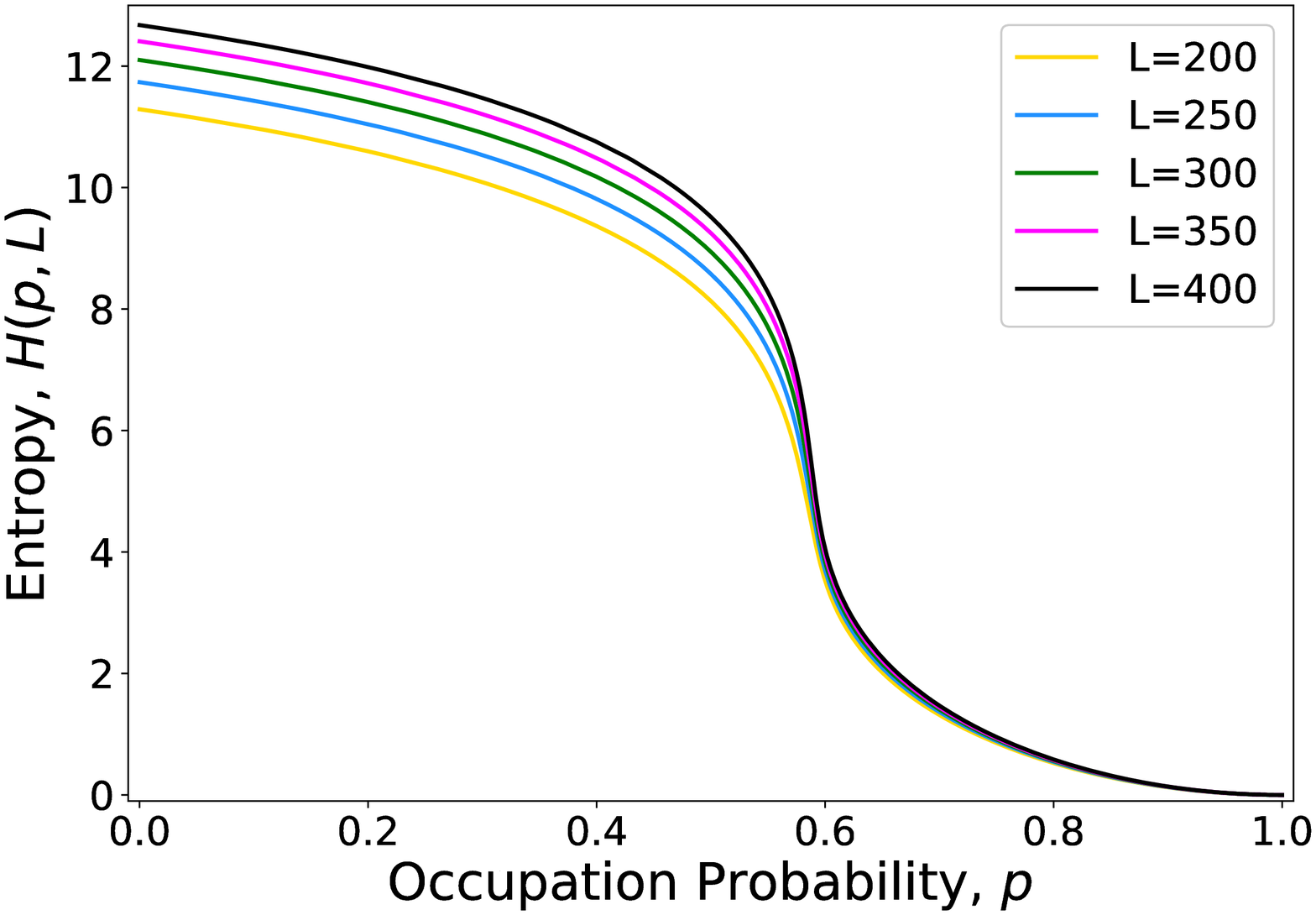}
\label{fig:2a}
}
\subfloat[]
{
\includegraphics[height=3.6 cm, width=4.0 cm, clip=true]
{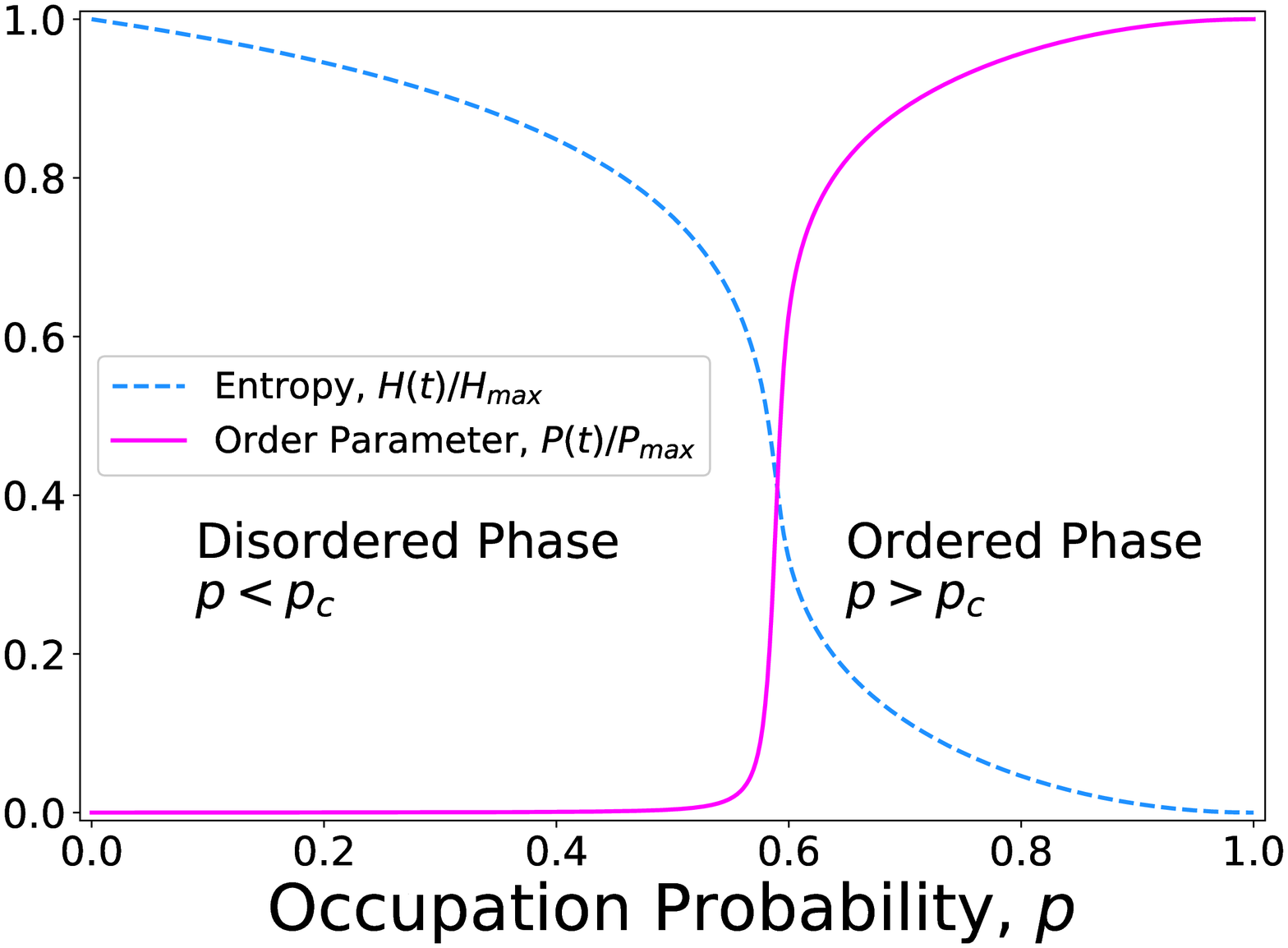}
\label{fig:2b}
}

\caption{(a) Plots of entropy $H$ versus $p$ for redefined site percolation. Its qualitative
behavior is now the same as for its bond counterpart. (b) Here we plot entropy $H(p)/H_{{\rm max}}$ and 
order parameter $P(p)/P_{{\rm max}}$
in the same graph to see the contrast.  It can be easily seen that $P=0$ where
entropy is maximally high and order parameter is maximally high where entropy is minimally low
which is reminiscent of order-disorder transition in the ferromagnetic transition.
} 

\label{fig:bond_site_ab}
\end{figure}

Using the redefined site percolation, we again measure entropy and find that it behaves just like
its bond counterpart. That is, entropy is 
maximum at $p=0$ and as $p$ approaches $p_c$ it starts dropping sharply where the occupation probability $p$ is 
now defined as  the fraction of the bonds being occupied. At above $p_c$ it then decreases slowly
to zero as $p\rightarrow 1$ which is shown in Fig. (\ref{fig:2a}) whose qualitative behaviour
is exactly the same as for the bond percolation shown in  Fig. (\ref{fig:1a}). 
Perhaps plots of entropy and order parameter
in the same graph can help us appreciate their opposing nature better than they are shown separately. 
To do that we re-scale entropy and  order parameter 
so that in either cases their respective maximum value is one. 
The plots of re-scaled entropy $H(p)/H(0)$  and order parameter $P(p)/P(1)$ as a function of $p$
are shown  in Fig. (\ref{fig:2b}).  It clearly shows that $H$ is maximally high where $P=0$ and 
the order parameter is maximally high where $H$ is minimally low. Moreover, we find
entropy never decreases with $1-p$ rather it either increases or remain constant and hence 
both the issues are resolved.

\begin{figure}

\centering

\subfloat[]
{
\includegraphics[height=3.6 cm, width=4.0 cm, clip=true]
{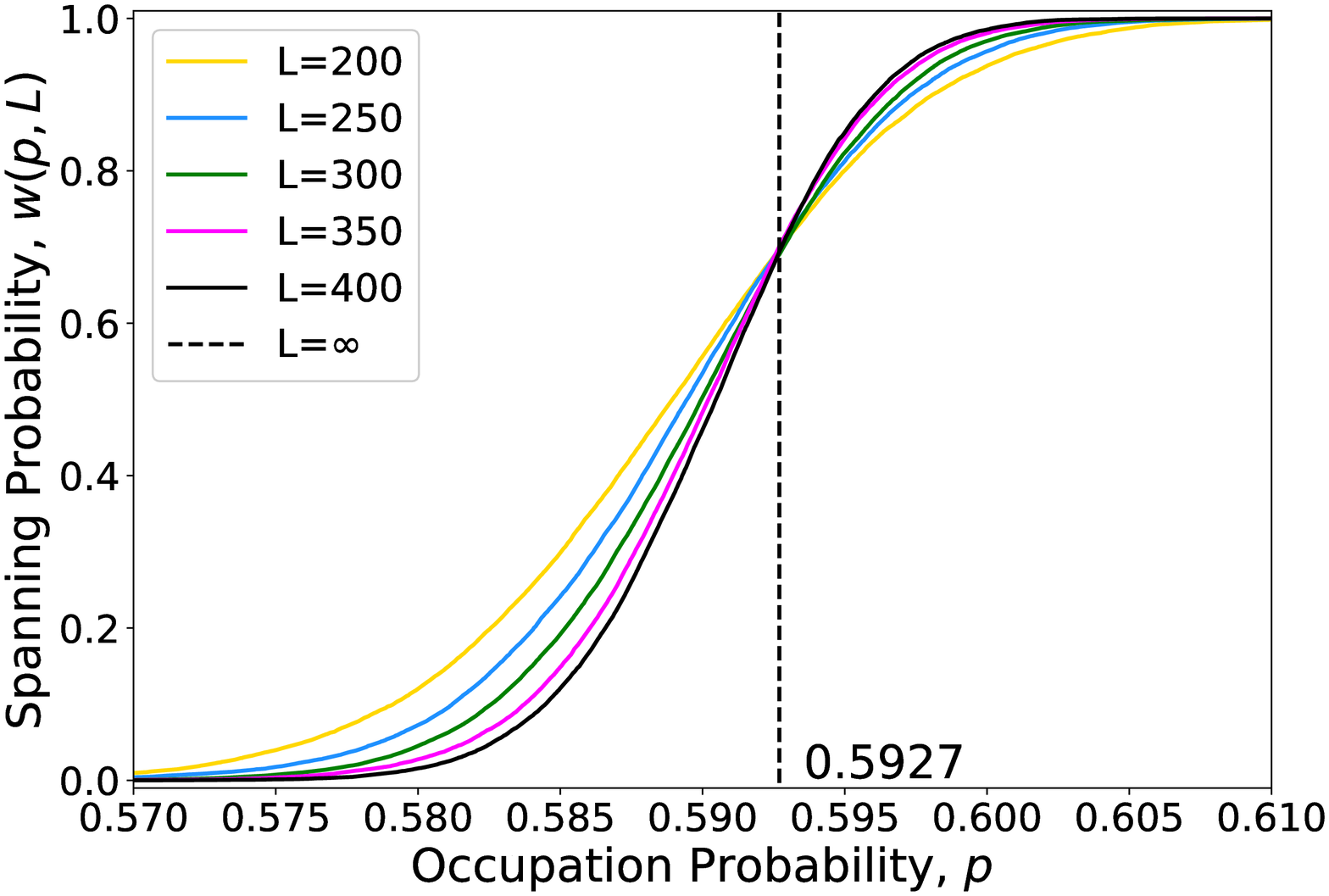}
\label{fig:3a}
}
\subfloat[]
{
\includegraphics[height=3.6 cm, width=4.0 cm,clip=true]
{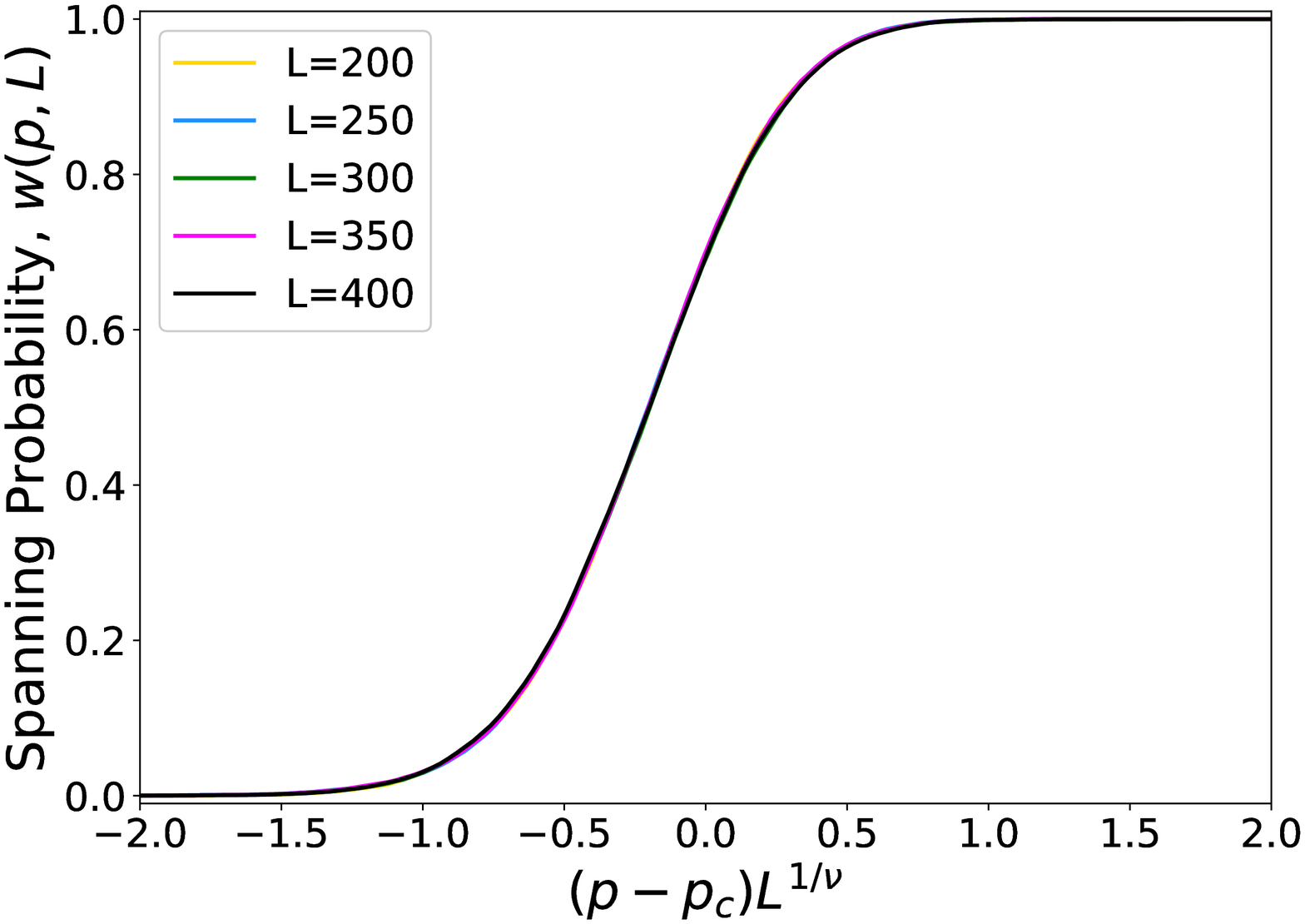}
\label{fig:3b}
}
\caption{(a) Spanning probability $W(p,L)$ vs $p$ for different lattice sizes using new definition
of site percolation. 
In (b) we plot dimensionless quantities $W$ vs $(p-p_c)L^{1/\nu}$ using known value of $\nu=4/3$
and find excellent data-collapse which is a proof that bond-site still belong to the same universality class.
} 

\label{fig:3ab}
\end{figure}

\section{Is site-bond universality still valid?}

We now check if the re-defined site percolation still gives the
same critical point $p_c=0.5927$ and the same critical exponent $\nu$ of the 
correlation length or not.
The best quantity for finding them is the
spanning probability $W(p)$. It describes the likelihood of finding a 
cluster that spans across the entire system
either horizontally or vertically at a given occupation probability $p$.
In Fig. (\ref{fig:3a}), we show $W(p)$ as a function of
$p$ for different system sizes $L$. One of the significant features of such plots is that they all 
meet at one particular value. It is actually the critical point $p_c=0.5927$
which is exactly the same known value as for the traditional definition of site percolation
despite it differs from our definition of site percolation.  
Thus, the value of $p_c$ does not depend on whether we measure the cluster size
in terms of the number of sites or the number of bond it contains.
Note that finding the $p_c$ value for different skeletons
is one of the central problems in percolation theory \citep{ref.ziff_pc_1, ref.ziff_pc_2}. 
To check whether the $\nu$ value is still the same we use its standard known value
$\nu=4/3$ in the finite-size scaling ansatz
\begin{equation}
W(p,L)\sim L^a\phi_w((p_c- p)L^{{{1}\over{\nu}}}),
\end{equation}
where $\phi_w$ is the universal scaling function for spanning probability \cite{ref.ziff_nu}. We then plot of $W(p)L^{-a}$ vs $(p_c- p)L^{{{1}\over{\nu}}}$ and
find that all the distinct curves of Fig. (\ref{fig:3a}) collapse into a universal 
scaling curve as shown in Fig. (\ref{fig:3b}) for $\nu=4/3$ with $a=0$. This is a clear testament
that the critical point $\nu$ is also the same as that of the traditional site percolation and that $W(p)$ behaves like a step function in the thermodynamic limit.

\begin{figure}

\centering

\subfloat[]
{
\includegraphics[height=3.6 cm, width=4.0 cm, clip=true]
{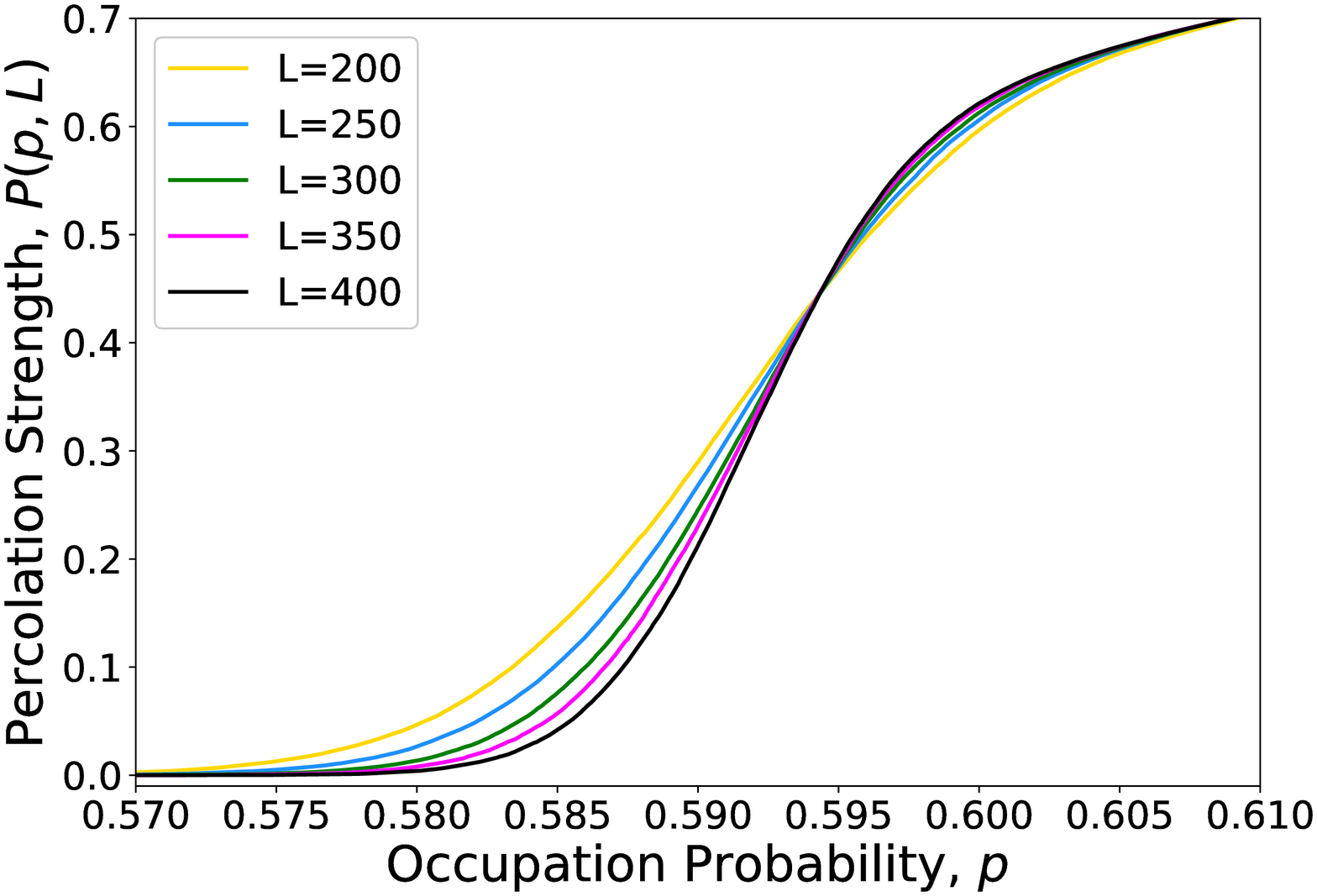}
\label{fig:4a}
}
\subfloat[]
{
\includegraphics[height=3.6 cm, width=4.0 cm, clip=true]
{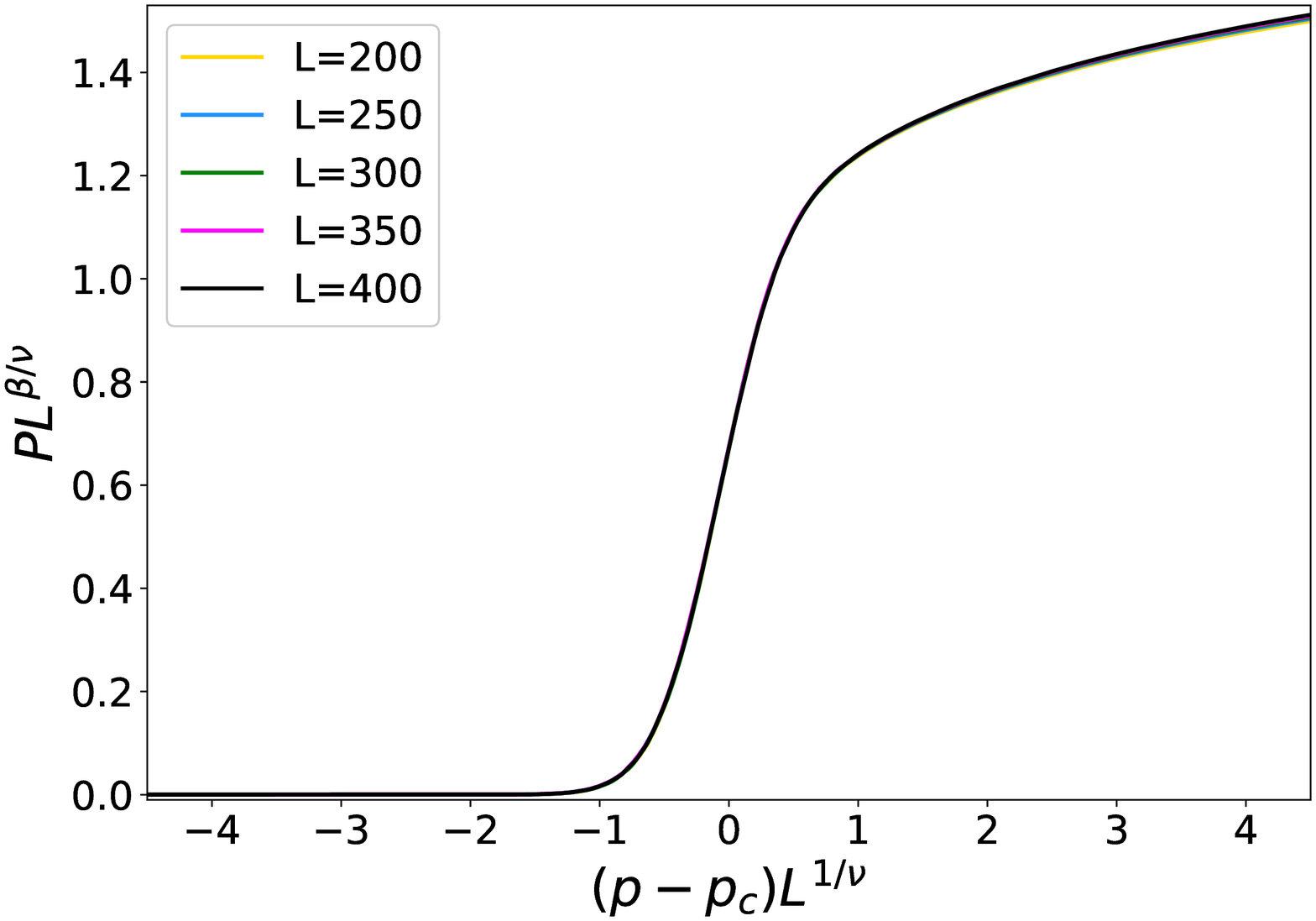}
\label{fig:4b}
}
\caption{(a) Order parameter $P(p,L)$ vs $p$ for re-defined site percolation in the square lattice. 
(b) We plot $P(p,L)L^{\beta/\nu}$ versus $(p-p_c)L^{1/\nu}$ using know value of $\nu=4/3$ and $\beta=5/36$. 
An excellent data collapse proves that our way defining site percolation can still reproduce the
same critical exponents. 
} 

\label{fig:4ab}
\end{figure}

Next we attempt to find the critical exponent $\beta$ of the order parameter $P$ using 
the new definition of site percolation.
First, we plot order parameter $P(p)$ in Fig. (\ref{fig:4a})  as a function of $p$ for different
lattice size $L$. We now apply the finite-size scaling 
\begin{equation}
\label{eq:fss_P}
P(p,L)\sim  L^{-\beta/\nu}\phi_p((p-p_c)L^{1/\nu}),
\end{equation}
where $\phi_p$ is the universal scaling function for order parameter. 
We now use the
standard known values for $\nu=4/3$ and $\beta/\nu=0.104$ to check if the plot 
of $P(p,L)L^{\beta/\nu}$ versus
$(p-p_c)L^{1/\nu}$ give data collapse or not. Indeed,
we find that all the distinct plots of Fig. (\ref{fig:4a}) collapse into
a universal scaling curve as shown in Fig. (\ref{fig:4b}). It implies
that the new definition of site percolation reproduces the same known value of $\beta=0.1388$ 
in $2d$ random percolation. Note that $P(p,L)L^{\beta/\nu}$  and $(p-p_c)L^{1/\nu}$ are dimensionless quantities and hence 
we have $P(p,L)\sim L^{-\beta/\nu}$ and $(p-p_c)\sim L^{-1/\nu}$. Eliminating $L$ from these relations we find
\begin{equation}
P\sim (p-p_c)^\beta.
\end{equation}
This is exactly how the order parameter, such as magnetization in the paramagnetic to 
ferromagnetic transition, behaves near the critical point.

\begin{figure}

\centering

\subfloat[]
{
\includegraphics[height=3.6 cm, width=4.0 cm, clip=true]
{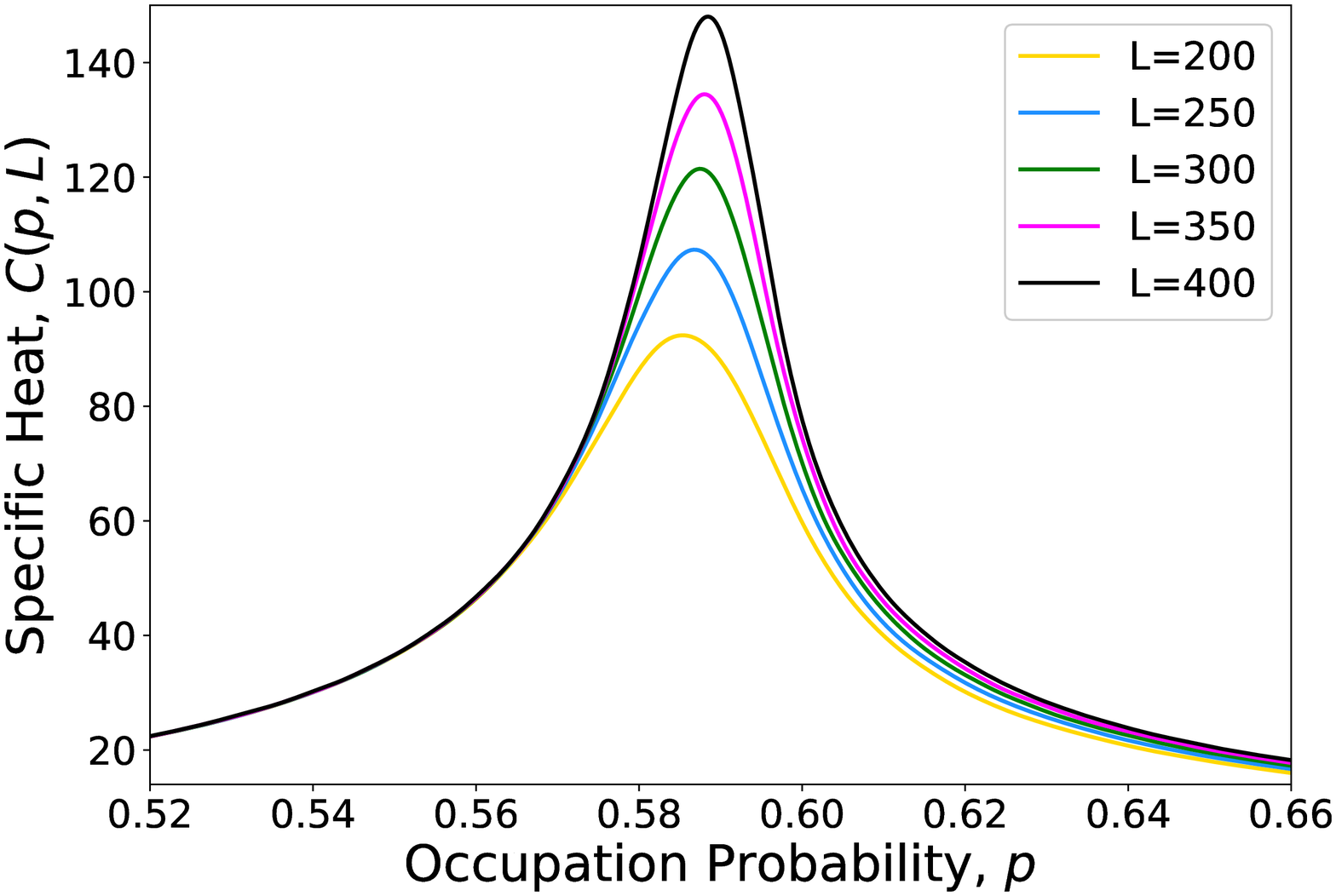}
\label{fig:5a}
}
\subfloat[]
{
\includegraphics[height=3.6 cm, width=4.0 cm, clip=true]
{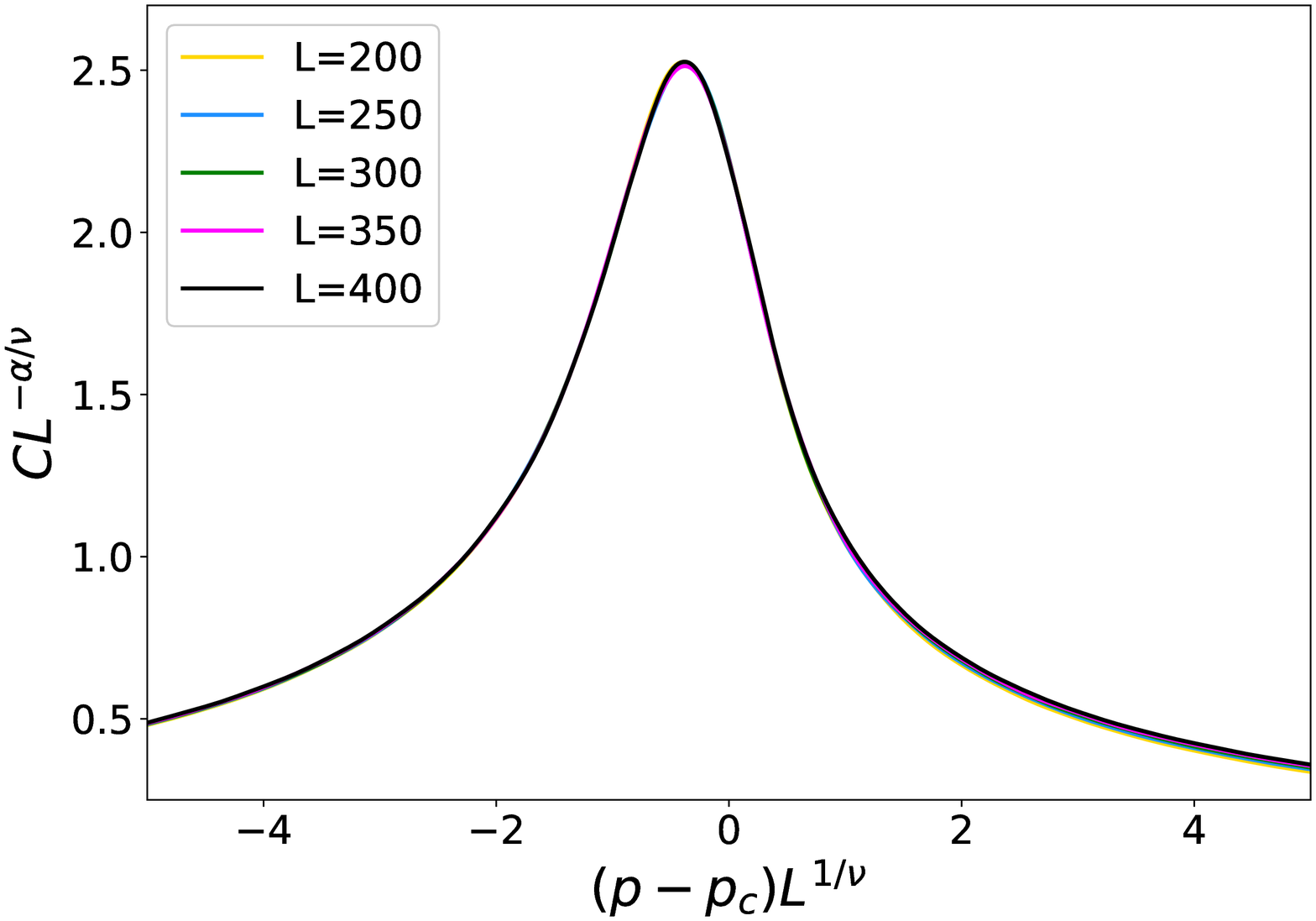}
\label{fig:5b}
}
\caption{Specific heat $C(p,L)$ vs $p$ in square lattice for re-defined site percolation. 
 In (b) we plot dimensionless quantities $CL^{-\alpha/\nu}$ vs $(p-p_c)L^{1/\nu}$ and we find an excellent data-collapse.
} 

\label{fig:5ab}
\end{figure}

 Knowing entropy paves the way for obtaining the
specific heat since we know that it is proportional to the first derivative of entropy
i.e. $C=TdS/dT$ where $S$ is the thermal entropy. If we now know the exact equivalent 
counterpart of temperature
then we can immediately obtain the specific heat for percolation. In our recent work we
argued that $1-p$ is the equivalent counterpart of temperature and hence the specific
heat for percolation is 
\begin{equation}
\label{eq:fss_c}
C(p)=(1-p){{dH}\over{d(1-p)}}.
\end{equation}
 The plots of $C(p)$ as a function of $p$ for different
system sizes $L$ is shown in Fig. (\ref{fig:5a}). 
Let us assume  that it obeys the finite-size scaling
\begin{equation}
C(p,L)\sim L^{\alpha/\nu}\phi_c((p-p_c)L^{1/\nu}),
\end{equation}
where $\phi_C$ is the scaling function. We already know the value of 
$\alpha=0.906$ from our recent work on bond percolation in
the square lattice \cite{ref.hassan_didar}. Using the same values for $\alpha$ and $\nu$ 
we now plot $CL^{-\alpha/\nu}$ vs $(p-p_c)L^{1/\nu}$ and find an excellent 
data collapse as shown in Fig. (\ref{fig:5b}).  It confirms that $\alpha= 0.906$ is indeed the
same for both bond and redefined site percolation. According to Eq. (\ref{eq:fss_c}) we have
$C\sim L^{\alpha/\nu}$ at $p=p_c$. Once again eliminating $L$ from it using $L\sim (p-p_c)^{-\nu}$ 
we find 
\begin{equation}
C\sim (p-p_c)^{-\alpha},
\end{equation}
which implies that the specific heat diverges at the critical point. This is one of the 
important attributes of the second order phase transitions. Divergence means that the thermodynamic 
property suffers a discontinuity and changes drastically from its value in the disordered state. 
It is qualitatively related to the emergence of long range order. 
The specific heat has so far been defined as the second derivative of the number of clusters $n(p)$ with
respect to $p$. Using this definition one find $\alpha = -2/3$ which is in sharp contrast to our result $\alpha=0.906$. It means specific
according to old definition does not diverges rather it shows a cusp. The old definition of specific heat also suggest that $n(p)$ is the free energy and 
hence its first derivative should have been the entropy. However, one do not get
expected entropy from this.

\begin{figure}

\centering

\subfloat[]
{
\includegraphics[height=3.6 cm, width=4.0 cm, clip=true]
{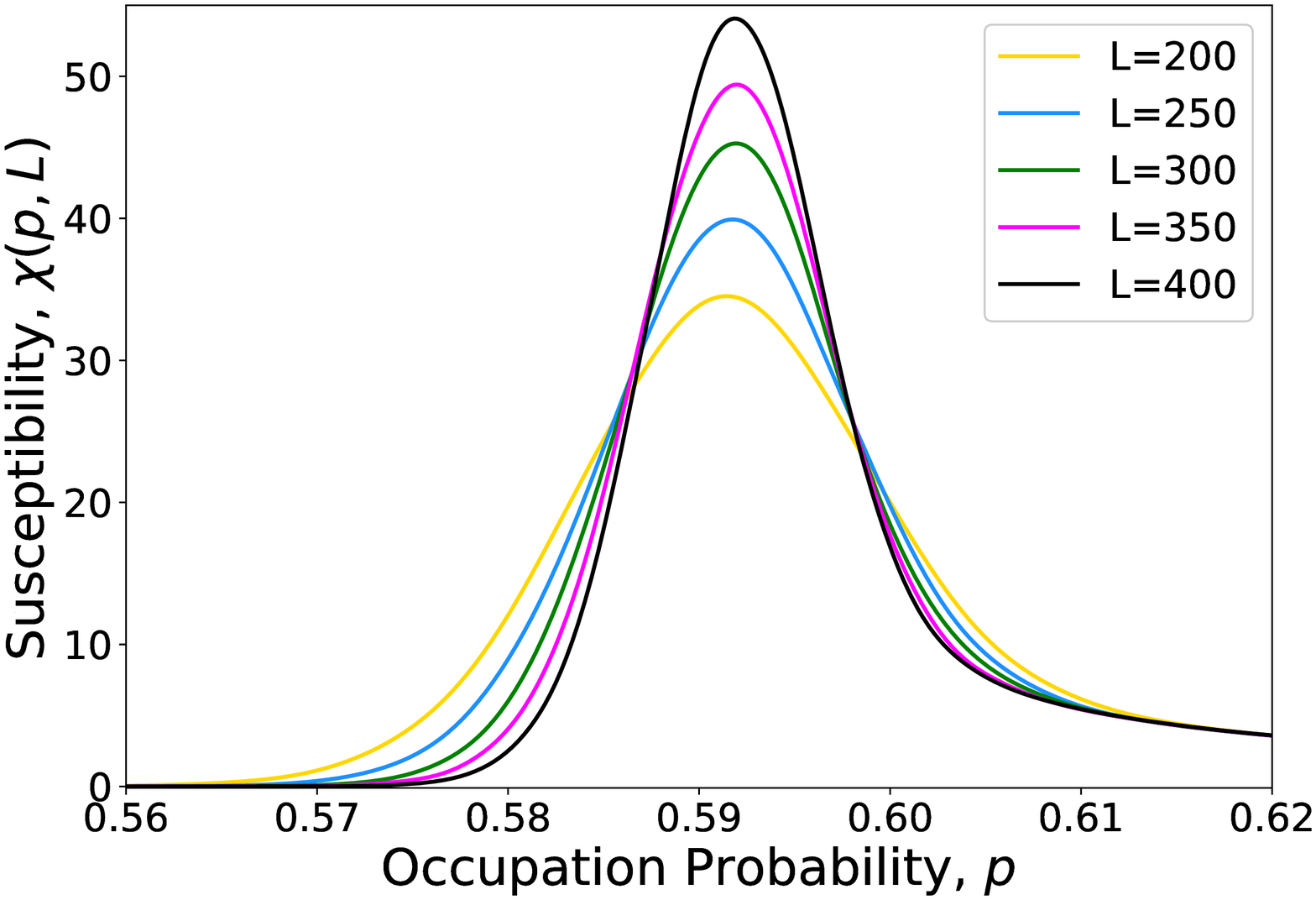}
\label{fig:6a}
}
\subfloat[]
{
\includegraphics[height=3.6 cm, width=4.0 cm, clip=true]
{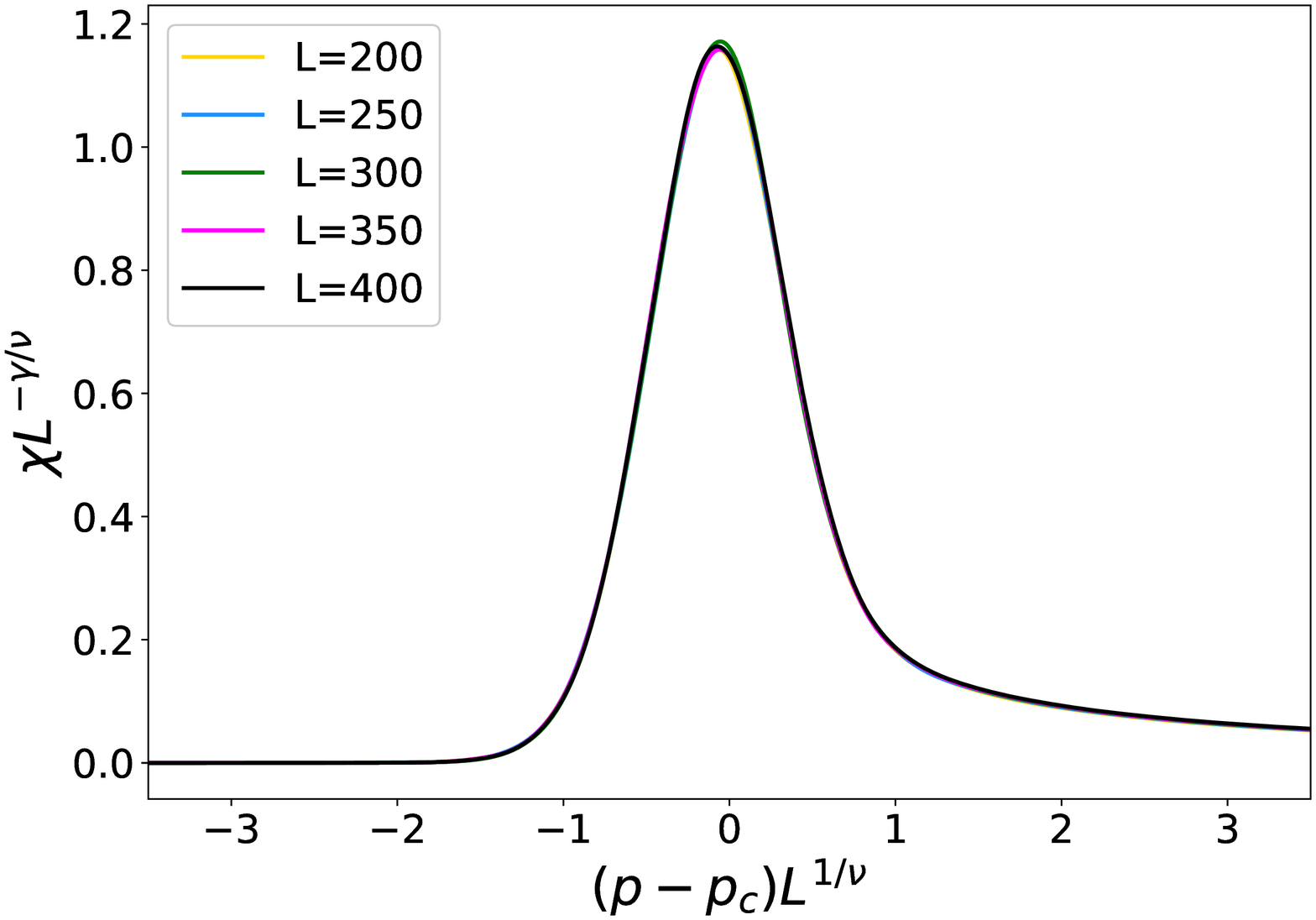}
\label{fig:6b}
}
\caption{(a) Plots of susceptibility $\chi(p)$ for redefined site percolation as a
function of $p$ in square lattice of different sizes. 
 In (b) we plot dimensionless quantities $\chi L^{-\gamma/\nu}$ vs $(p-p_c)L^{1/\nu}$ and we find an excellent data-collapse with $\gamma=0.853$ which is the same as for bond type.
} 

\label{fig:6ab}
\end{figure}

In percolation, yet another quantity of interest is the susceptibility. 
Traditionally, mean cluster size
has been regarded as the equivalent counterpart of susceptibility. Sometimes 
variance of the order parameter $\langle P^2\rangle -\langle P\rangle^2$ too is
regarded as susceptibility. Neither of the two actually gives respectable value for $\gamma$ to obey the
Rusbrooke inequality. Recently, we have proposed susceptibility $\chi(p,L)$ for percolation as
the ratio of the change
in the order parameter $\Delta P$  and the magnitude of the time interval $\Delta p$ during which 
the change $\Delta P$  occurs i.e.,
\begin{equation}
\chi(p,L)={{\Delta P}\over{\Delta p}}.
\end{equation}. 
Essentially it becomes the derivative of the order parameter $P$ since
$\Delta p\rightarrow 0$ in the limit  $N\rightarrow \infty$ as $\Delta p={{1}\over{2L^2}}$. The idea
of jump has been studied first by Manna in the context of explosive percolation \cite{ref.manna}.
The resulting susceptibility is shown
in Fig. (\ref{fig:6a}) as a function of $p$. We already know that $\gamma/\nu=0.6407$ for 
bond percolation in the square lattice. Using the same value for redefined
site percolation in the plot of $\chi L^{-\gamma/\nu}$ vs $(p-p_c)L^{1/\nu}$ 
we find that all the distinct curves in Fig. (\ref{fig:6a}) collapse superbly. It confirms
that the susceptibility obeys the finite-size scaling
\begin{equation}
\chi\sim L^{\gamma/\nu}\phi_\chi((p-p_c)L^{1/\nu}).
\end{equation}
It once again suggest that susceptibility diverges near the critical point following a power-law
\begin{equation}
\chi\sim (p-p_c)^{-\gamma},
\end{equation}
with $\gamma=0.8542$. It implies
once again that bond and redefined site percolation share the same $\gamma$ value.

\section{Conclusions}

In this article we first discussed entropy for bond percolation
and we measured it as a function of $p$. We have found that it is consistent with the behaviour 
of the order parameter and with the second law of thermodynamics. Essentially, entropy measures 
the degree of disorder while order parameter measures the extent of order. Thus, both 
cannot be minimum or maximum 
at the same state since the system cannot be in the most disordered and in the 
most ordered state at the same time.
We have then measured entropy and order parameter for site percolation using its existing definition.
We have found 
that at $p=0$ both order parameter and entropy equal 
to zero, which is contradictory. Besides, we have found that entropy first increases with $1-p$ and then
decreases again. This is a clear violation of the fact entropy can either increase or
remain constant with temperature but cannot decrease.  
We have therefore redefined the site percolation as follows. We occupy sites to connect 
bonds which are assumed to exist already in the system and measure cluster sizes
in terms of the number of contiguous bonds connected by occupied sites. 

With the new definition we have found that the entropy behaves exactly in the same way as 
it does in the case of its bond counterpart. Thus the conflict that the system is in ordered and disordered 
state at the same time is resolved and it obeys the second law of thermodynamics too. 
 One of the most important well-known results is that site and bond type percolation in a  given $d$
dimensional lattice belong to the same universality class regardless of the detailed nature of the 
structure of the lattice. We have shown that this is still true despite 
the old and new definitions are distinctly different. It proves that changing local rules on a regular lattice does not
change the universality class of the system.

Note that occupation of one isolated site forms a cluster of size four according to new definition while according to old definition it forms
 a cluster of size one only. In this sense the bond percolation is also different as
 we find that occupation of one isolated bond forms a cluster of size two.  
 We hope the present work will have significant impact in the future research of percolation
theory.

\end{document}